\documentclass[aip,jcp,amsmath,amssymb,floatfix,reprint,citeautoscript]{revtex4-2}

\usepackage[utf8]{inputenc}
\usepackage{mathptmx}
\usepackage{graphicx}
\usepackage{wrapfig}
\usepackage[sectionbib]{bibunits}
\usepackage[colorlinks,allcolors=black,citecolor=blue,urlcolor=blue]{hyperref}
\usepackage{booktabs} 
\usepackage{silence}
\usepackage{multirow}

\begin{document}

\def\mstitle{Common Cations are not Polarizable: Effects of Dispersion Correction on Hydration Structures from Ab Initio Molecular Dynamics}
\title{\mstitle}

\author{Vojtech Kostal}

\author{Philip E. Mason}

\author{Hector Martinez-Seara*}
\email{hseara@gmail.com}

\author{Pavel Jungwirth*}
\email{pavel.jungwirth@uochb.cas.cz}

\affiliation{
Institute of Organic Chemistry and Biochemistry of the Czech Academy of Sciences, Flemingovo nám. 2, 166 10 Prague 6, Czech Republic
}

\date{\today}

\begin{abstract}

\setlength\intextsep{0pt}
\begin{wrapfigure}{r}{0.4\textwidth}
  \hspace{-1.8cm}
  \includegraphics[width=0.4\textwidth]{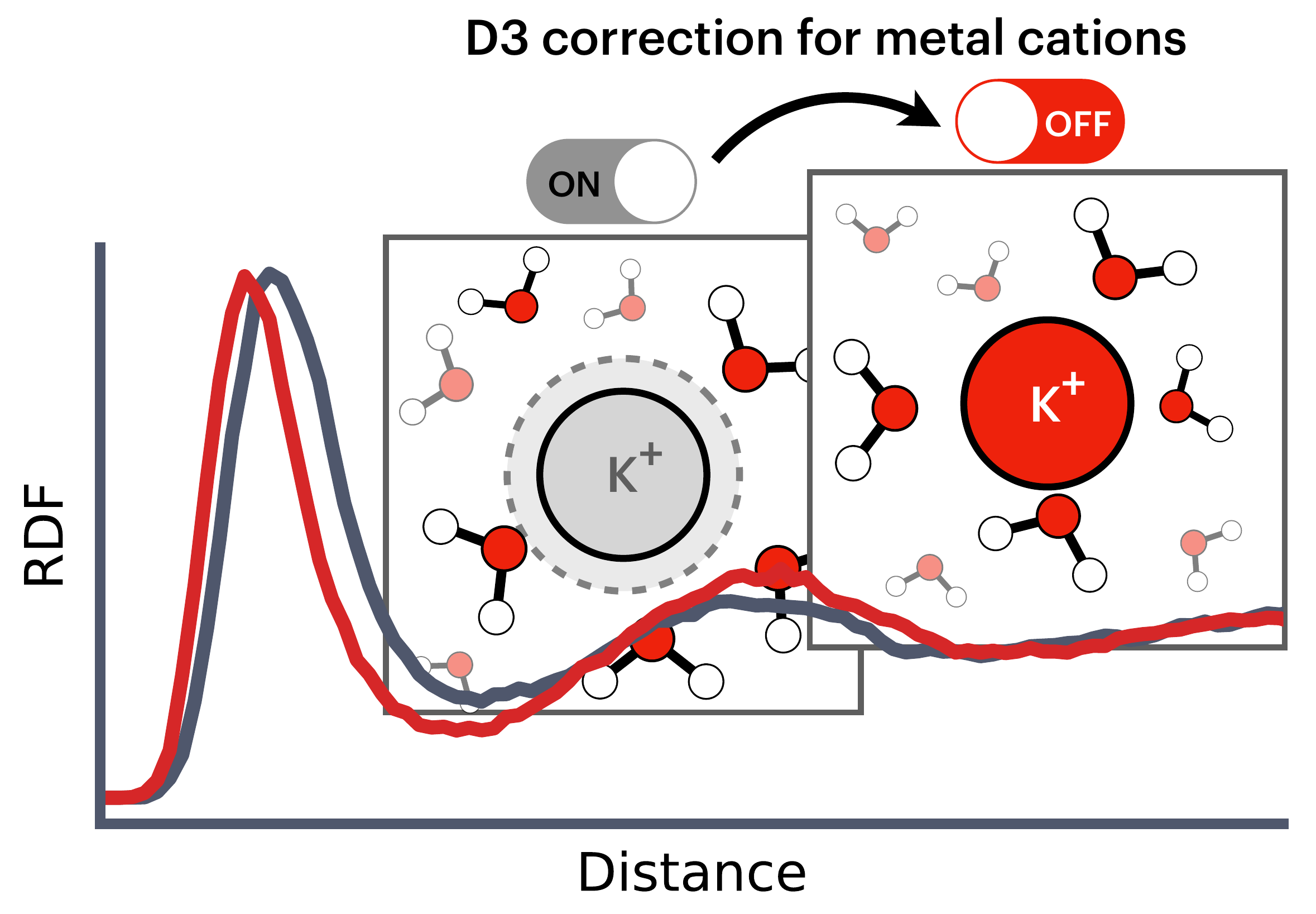}
\end{wrapfigure}

We employed density functional theory-based ab initio molecular dynamics simulations to examine the hydration structure of several common alkali and alkali earth metal cations.
We found that the commonly used atom pairwise dispersion correction scheme D3, which assigns dispersion coefficients based on the neutral form of the atom rather than its actual oxidation state, leads to inaccuracies in the hydration structures of these cations.
We evaluated this effect for lithium, sodium, potassium, and calcium and found that the inaccuracies are particularly pronounced for sodium and potassium compared to the experiment.
To remedy this issue, we propose disabling the D3 correction specifically for all cation-including pairs, which leads to a much better agreement with experimental data.

\end{abstract}

{\maketitle}

\begin{bibunit}

Dispersion interaction can be important even in systems where electrostatic forces dominate.
A prominent example is liquid water, where lack of dispersion in common density functional theory (DFT) methods leads to a qualitatively flawed description~\cite{Schmidt2009-10.1021/JP901990U, Lin2012-10.1021/CT3001848, Gillan2016-10.1063/1.4944633}.
Indeed, ab initio molecular dynamics (AIMD) simulations with standard generalized gradient approximation (GGA) functionals predict water to be solid at room temperature~\cite{Yoo2011-10.1063/1.3573375, Morawietz2016-10.1073/PNAS.1602375113}.
This problem can be largely fixed by adding an empirical dispersion term (D)~\cite{Grimme2004-10.1002/JCC.20078, Grimme2016-10.1021/ACS.CHEMREV.5B00533}.

Within the generally employed D3 scheme~\cite{Grimme2010-10.1063/1.3382344}, atom-specific London dispersion coefficients are assigned independent of the electronic density.
This works well in most cases. However, one should be careful when large changes in electronic density occur, such as when moving from a neutral atom to the corresponding cation.
For example, a sodium atom has a polarizability of 24~\AA$^3$~\cite{Ekstrom1995-10.1103/PhysRevA.51.3883}, while that of the sodium cation is two orders of magnitude smaller, amounting to only 0.18~\AA$^3$~\cite{Molina2011-10.1063/1.3518101}.
Modeling aqueous alkali halide salt solutions using atomic rather than ionic polarizabilities may thus lead to severe artifacts.
Note that this approach is widespread in AIMD simulations despite the fact that the authors of the D3 method explicitly pointed the issue out.
Also note that the more recent variant of the dispersion correction D4~\cite{Caldeweyher2017-10.1063/1.4993215, Caldeweyher2019-10.1063/1.5090222} as well as the Tkatchenko--Scheffler method~\cite{Tkatchenko2012-10.1103/PHYSREVLETT.108.236402} take to a certain extent hybridization effects on the polarizabilities into account.
However, they still fall short of adequately accounting for the large differences between atomic and cationic polarizabilities.

In this study, we investigate the effect of applying atomic polarizabilities on the hydration structure of selected alkali and alkali earth cations in dilute and concentrated aqueous salt solutions.
We demonstrate that this effect is sizable and can, in some cases, like for sodium or potassium cations, lead to the qualitatively incorrect description of the ionic hydration shell and ion pairing.
We also suggest a simple and efficient fix: setting the cationic dispersion coefficients to zero.

\begin{figure}[h]
    \centering
    \includegraphics[width=\linewidth]{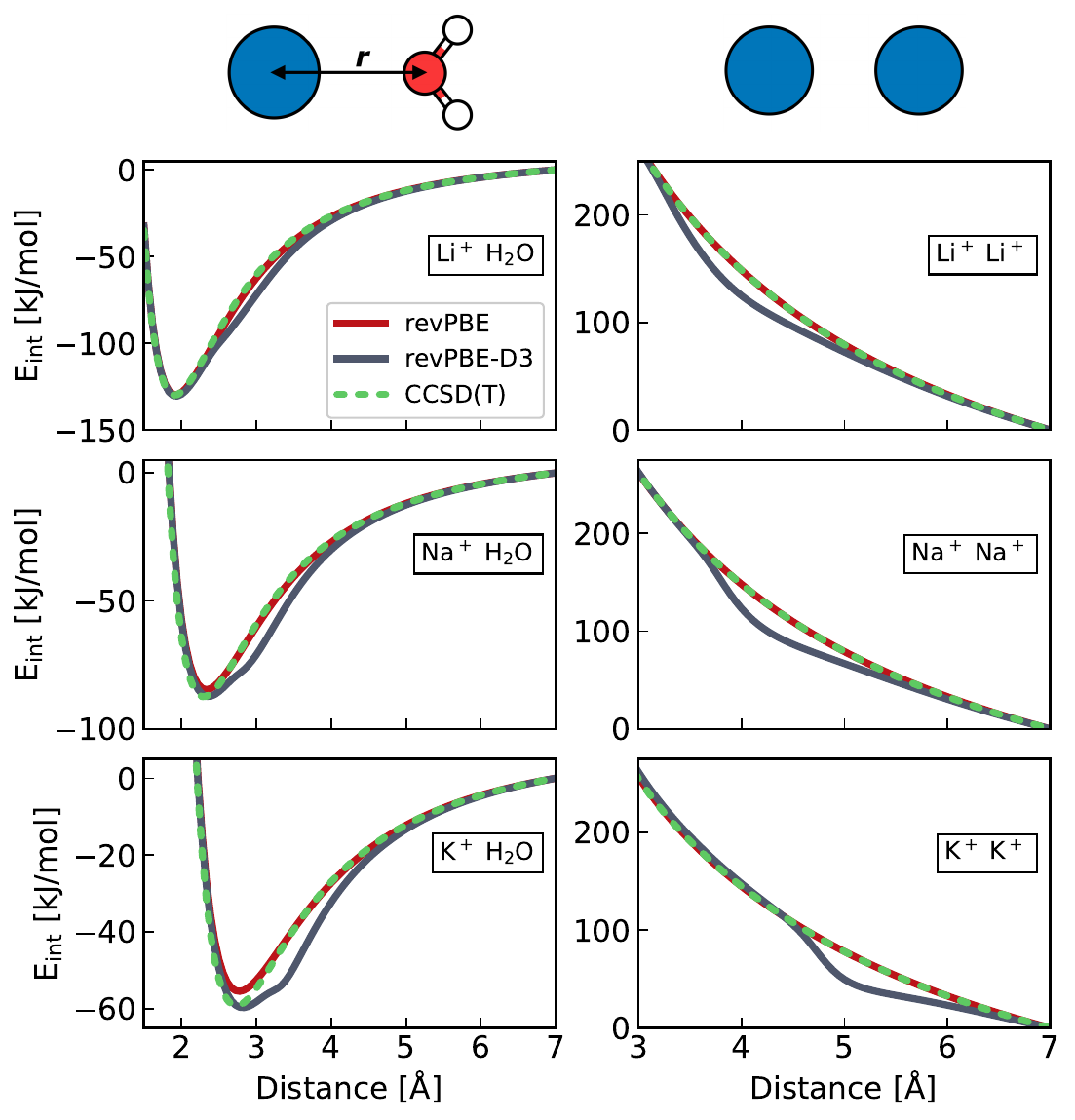}
    \caption
    {
    Interaction energy curves in the gas phase as a function of distance for cation--water (left) and cation--cation (right).
    The top panel illustrates the orientation of the water molecule with respect to the cation (blue) and the distance $r$ used in the interaction energy scan.
    The curves are color-coded to indicate the employed method: grey for revPBE-D3, red for revPBE, and dashed green for CCSD(T).
    Note that the $E_\mathrm{int}$ is aligned to zero at the largest distance for all cases.
    }
    \label{fig:Eint-gas-phase}
\end{figure}

In order to demonstrate for cations the artificial effect of the dispersion correction, we first present gas-phase interaction energies ($E_\mathrm{int}$) of cation--water and cation--cation pairs calculated using the revPBE and revPBE-D3 density functionals, as compared to the coupled cluster [CCSD(T)] method as the ``golden standard`` (Figure~\ref{fig:Eint-gas-phase}).
At large to intermediate separations, the interaction energies are determined primarily by charge--dipole and charge--charge electrostatic interactions in the case of cation-water and cation--cation pairs, respectively.
These interactions are well captured by DFT, hence the very good agreement between the revPBE and CCSD(T) potential energy curves.
Note that the revPBE method does not account for dispersion interactions while CCSD(T) does, confirming that dispersion is negligible for the cations investigated here.

The D3 correction incorporated in the revPBE-D3 functional
results for the investigated systems in a spurious shoulder on the potential energy curves (Figure~\ref{fig:Eint-gas-phase}).
This artificial stabilization of 10--30 kJ/mol, which is a direct consequence of the D3 parametrization treating in terms of dispersion atomic cations as neutral atoms, indeed occurs around the positions of hypothetical van der Waals minima between the corresponding neutral pairs.
Note that the D4 parameterization, which considers not only the atomic specificity but also the oxidation state, does not remove --- but only slightly reduces --- the size of this artifact (see SI for more details).
Also note that in relative terms, the artifact is larger for sodium and potassium than lithium and calcium (see SI) due to weaker electrostatic interactions in the case of the former cations.
Finally, we stress that the purpose of the present gas phase calculations is to point directly to the artifact due to the inappropriate use of the atomic dispersion term for the corresponding cation, which is helpful for understanding the analogous effect on the structural properties of aqueous salt solutions in the condensed phase (\emph{vide infra}).

\begin{figure}[t!]
    \centering
    \includegraphics[width=\linewidth]{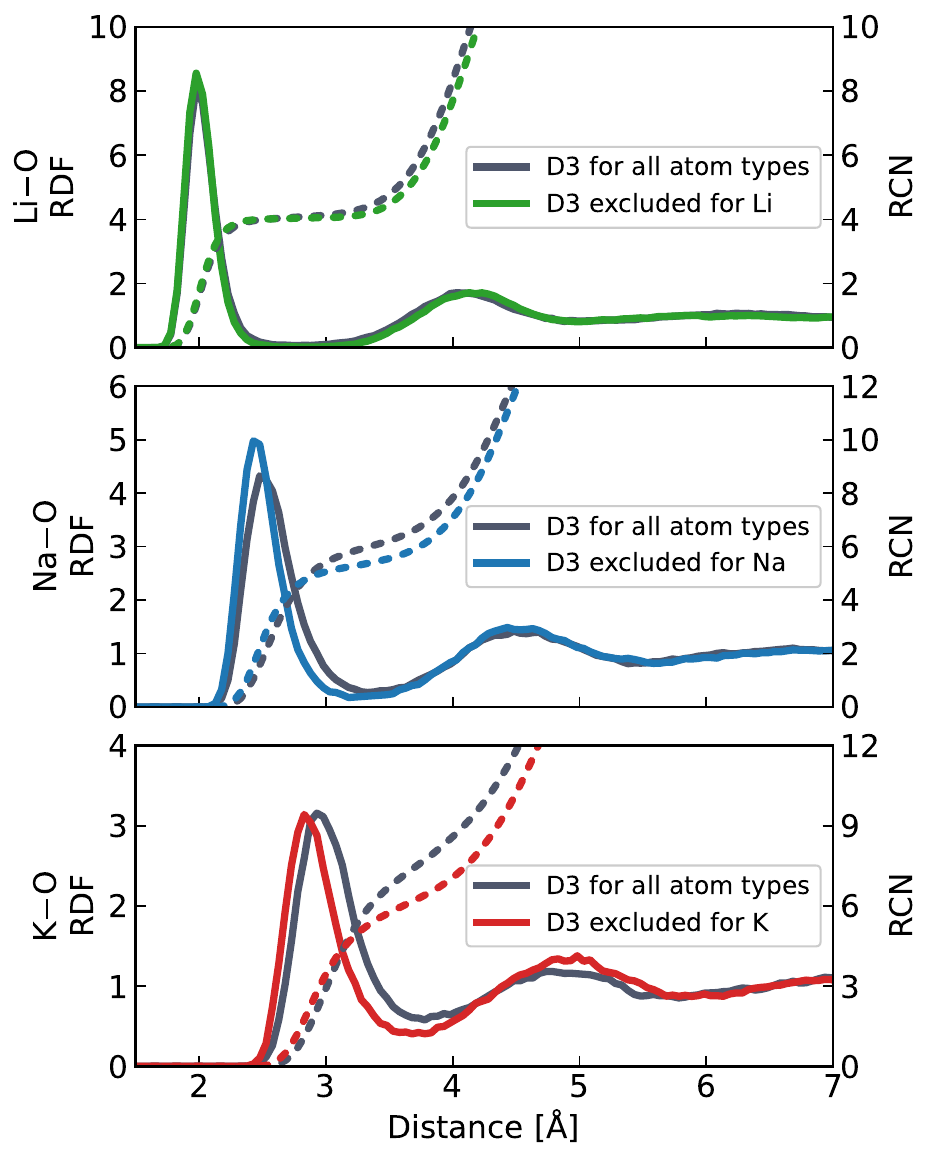}
    \caption
    {
    Radial distribution functions of cation--oxygen (full) and their corresponding running coordination numbers (dashed) including the D3 correction for all atoms (grey); and all atoms except Li (green), Na (blue), and K (red) cations.
    The panels are arranged in the order of lithium (top), sodium (middle),  and potassium (bottom).
    }
    \label{fig:rdf-NaOW-LiOW}
\end{figure}

\begin{figure*}[t!]
    \centering
    \includegraphics[width=0.85\linewidth]{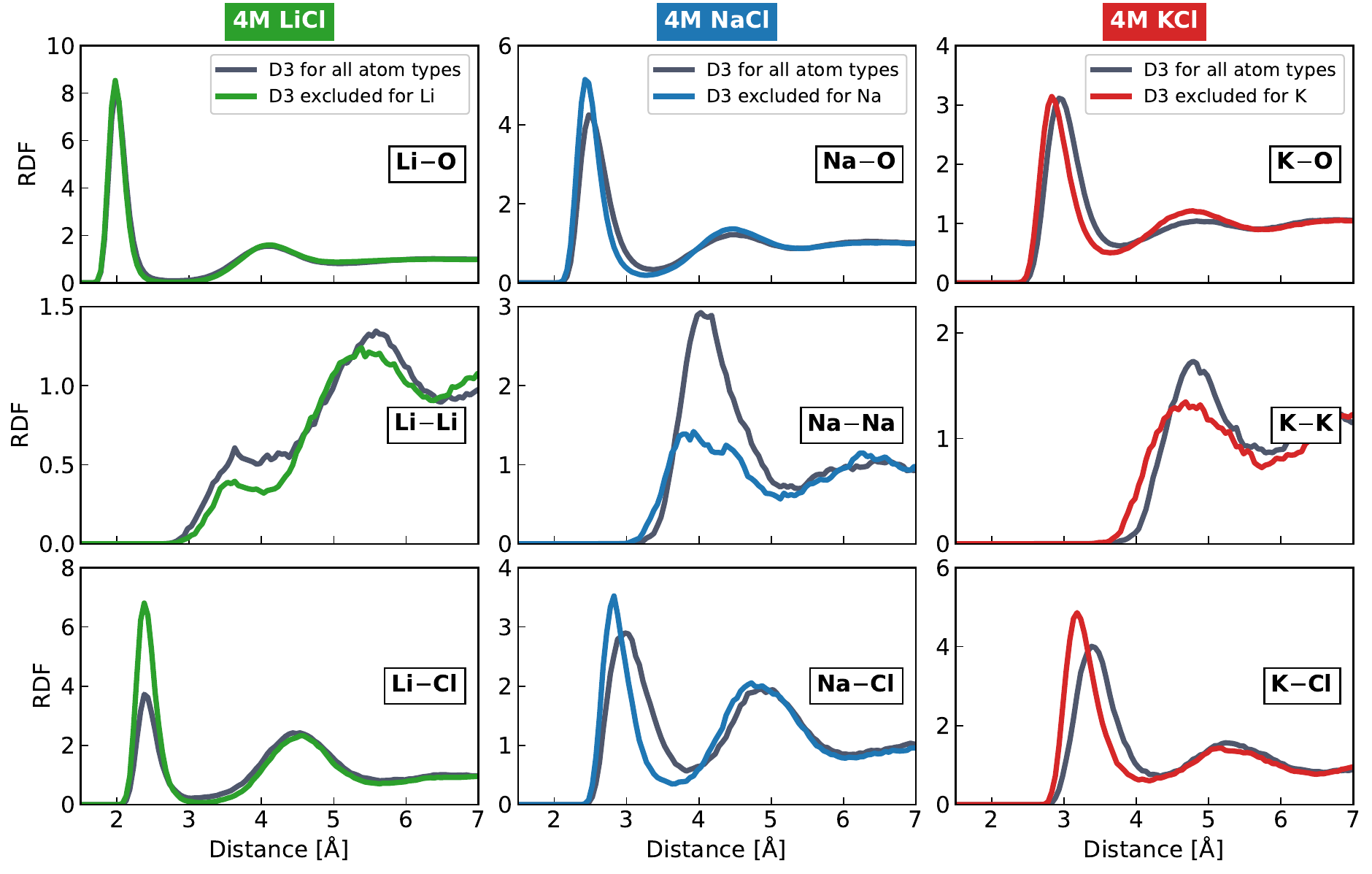}
    \caption
    {
    Radial distribution functions of metal cations in 4~M LiCl (left), 4~M NaCl (middle), and 4~M KCl (right) solutions.
    The top, middle, and bottom panels show the radial distribution functions of water oxygens, metal cations, and chloride anions, respectively, surrounding metal cations.
    The results are shown with the D3 correction turned on for all atom types in grey and for all types except Li (green), Na (blue), and K (red) cation.
    }
    \label{fig:rdf-4M}
\end{figure*}

Two sets of AIMD simulations of a single aqueous Li$^+$, Na$^+$, K$^+$ or Ca$^{2+}$ (see SI) cations were performed: one with the D3 dispersion correction enabled for all atoms (\textit{i.e.}, default setting) and second with the D3 term disabled for all atomic pairs involving the cation.
The hydration structure of the cations was quantified using radial distribution functions (RDFs) and running coordination numbers (RCNs) of the surrounding water molecules, as presented in Figure~\ref{fig:rdf-NaOW-LiOW}.
We see sizable artifacts due to the D3 correction for sodium and potassium cations.
Namely, the presence of the dispersion term on these cations leads to a looser hydration shell which is shifted to a larger separation from the cation and contains about one additional water molecule.
The artificially large dispersion term thus effectively increases the interaction of the cation with the surrounding water molecules but at the same time reduces its ability to reorient and thus bring closer the neighboring water molecules.
This is because dispersion interactions depend much less on the ion-water molecule relative orientation than the electrostatic ion-dipole interactions.    
Comparison between the four investigated cations shows that the strength of the electrostatic interaction plays an important role.
In particular, for the two cations with a high charge density and thus stronger electrostatic interactions, \textit{i.e.}, for lithium and calcium, the relative importance of the dispersion interaction artifact is smaller than for sodium and potassium, which have a lower charge density (Figure~\ref{fig:rdf-NaOW-LiOW}).

Quantitatively, as detailed in Table~\ref{tab:rdfmax-cn-summary}, upon zeroing the cationic D3 terms, a shift of 0.05~\AA, 0.10~\AA\ in the position of the first peak of the Na--O and K--O RDF respectively is apparent as well as significant changes in the average coordination number (CN), moving from 6.04 to 5.32 for sodium and 7.70 to 6.39 for potassium.
It is reassuring that disabling the cationic D3 correction leads to better agreement with neutron scattering experiments for sodium and potassium (Table~\ref{tab:rdfmax-cn-summary}), in particular in terms of the average coordination numbers.
Note that accurate hydration structure was also achieved by the SCAN functional~\cite{Duignan2020-10.1039/C9CP06161D}, which bypasses the use of empirical dispersion being, however, less accurate for the structure of liquid water compared to revPBE-D3.

\begin{table}[b!]
    \centering
    \caption
    {
    Metal--water oxygen radial distribution function first peak positions (RDF$_\mathrm{max}$) and average coordination numbers (CN) for systems at infinite dilution (Figure~\ref{fig:rdf-NaOW-LiOW}) and at 4~M chloride solution (Figure~\ref{fig:rdf-4M}).
    Experimental values were adopted from Reference~\citenum{Pluharova2013-10.1021/JP402532E} for lithium, \citenum{Ansell2006-10.1016/j.bpc.2006.04.018} for sodium and, \citenum{Mason2019-10.1021/ACS.JPCB.9B08422} for potassium.
    The use of the cationic D3 correction is indicated in parentheses.
    }
    \begin{tabular}{lcccccc}
        \toprule
        & \multicolumn{2}{c}{Infinite Dilution} & \multicolumn{2}{c}{4~M solution} & \multicolumn{2}{c}{experiment} \\
        \cmidrule(r{1pt}){2-7}
        ion & RDF$_\mathrm{max}$ [\AA] & CN &  RDF$_\mathrm{max}$ [\AA] & CN & RDF$_\mathrm{max}$ [\AA] & CN \\
        \midrule
        Li$^+$ & 1.98 & 4.07 & 1.98 & 3.70 & \multirow{2}{*}{1.96} & \multirow{2}{*}{4} \\
        Li$^+$ (D3) & 1.98 & 4.15 & 1.98 & 3.93 &\\
        Na$^+$ & 2.43 & 5.32 & 2.43 & 5.19 & \multirow{2}{*}{2.38} & \multirow{2}{*}{5}\\
        Na$^+$ (D3) & 2.48 & 6.04 & 2.48 & 5.63 & \\
        K$^+$ & 2.83 & 6.39 & 2.83 & 5.73 & \multirow{2}{*}{2.96} & \multirow{2}{*}{6.1}\\
        K$^+$ (D3) & 2.93 & 7.70 & 2.93 & 7.03 & \\
    \end{tabular}
    \label{tab:rdfmax-cn-summary}
\end{table}

Next, we simulated concentrated (4~M) aqueous solutions of lithium, sodium, and potassium chloride with and without the cationic D3 correction which allowed us to acquire cation--water, cation--cation, and cation--anion RDFs.
These conditions are close to the experimental setup used in neutron scattering measurements, which are typically conducted at higher salt concentrations, compared to the infinite dilution.
The cation--water RDFs shown in the top part of Figure~\ref{fig:rdf-4M} display the same characteristics as those from the infinitely dilute systems, but with the CN of sodium cation closer to the experimental value found in Table~\ref{tab:rdfmax-cn-summary}.
In concentrated solutions, the second peak of the Na--O RDF is also affected by the D3 correction on sodium, in contrast to the infinitely diluted system.
However, the general trend of increased CNs for cations with the D3 correction still holds.

The D3 correction has a pronounced effect on sodium and potassium cations, while there are also non-negligible changes for the lithium cation.
Namely, the first peaks of the cation--water RDFs are larger when the D3 correction is applied to the cations. This increase is due to the presence of the shoulders at 3--4~\AA\ at the potential energy curves between cations and water, as well as at 4--5~\AA\ at the potential energy curves between a pair of cations (Figure~\ref{fig:Eint-gas-phase}).

Simulations of concentrated solutions provide insights also into cation--cation interactions, as well as interactions between cations and anions, as shown in the middle and bottom panels of Figure~\ref{fig:rdf-4M}.
Again, the peak intensities are larger when the D3 is applied to cations.
This is the most pronounced for sodium, and in the case of K--K RDF, there is also a noticeable shift in the position of the first peak.
Similar, although less noticeable, effects are also observed in the Li--Li RDF.

\begin{figure}[t!]
    \centering
    \includegraphics[width=\linewidth]{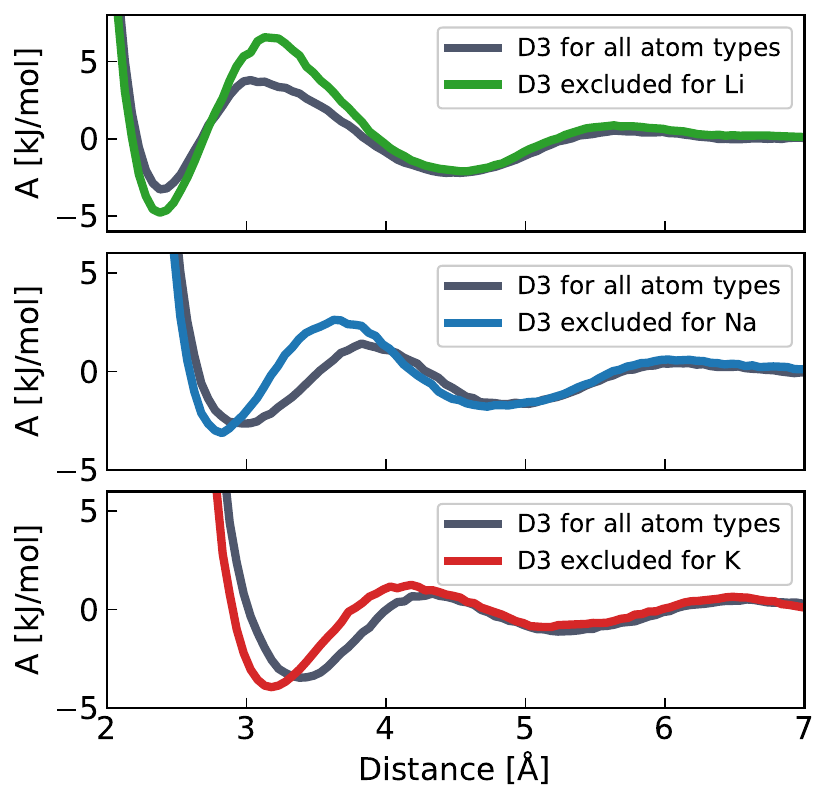}
    \caption{Free energy profiles of cation--chloride pairing as a function of distance for lithium chloride (top, green) and sodium chloride (middle, blue) and potassium chloride (bottom, red) obtained from the corresponding RDFs from Figure~\ref{fig:rdf-4M}.}
    \label{fig:pmf-nacl-licl}
\end{figure}

The cation--anion RDFs, shown in the bottom panels of Figure~\ref{fig:rdf-4M}, exhibit significant differences with or without dispersion.
The first peak is more pronounced when the D3 correction is turned off for metallic cations, and in the case of the Na--Cl and K--Cl RDFs and the position of the first maximum and first minimum are shifted by about 0.2~\AA, 0.3~\AA\ respectively.
These RDFs can be directly converted to the free energies of ion pairing (Equation~\ref{eq:free-energy}), as shown in Figure~\ref{fig:pmf-nacl-licl}.
The above observations have important implications for the parametrization of classical ionic force fields, where the AIMD-based free energies of ion binding have been used as a reference~\cite{Duboue-Dijon2018-10.1021/ACS.JPCB.7B09612, Martinek2018-10.1063/1.5006779}.

\begin{figure}[ht]
    \centering
    \includegraphics[width=\linewidth]{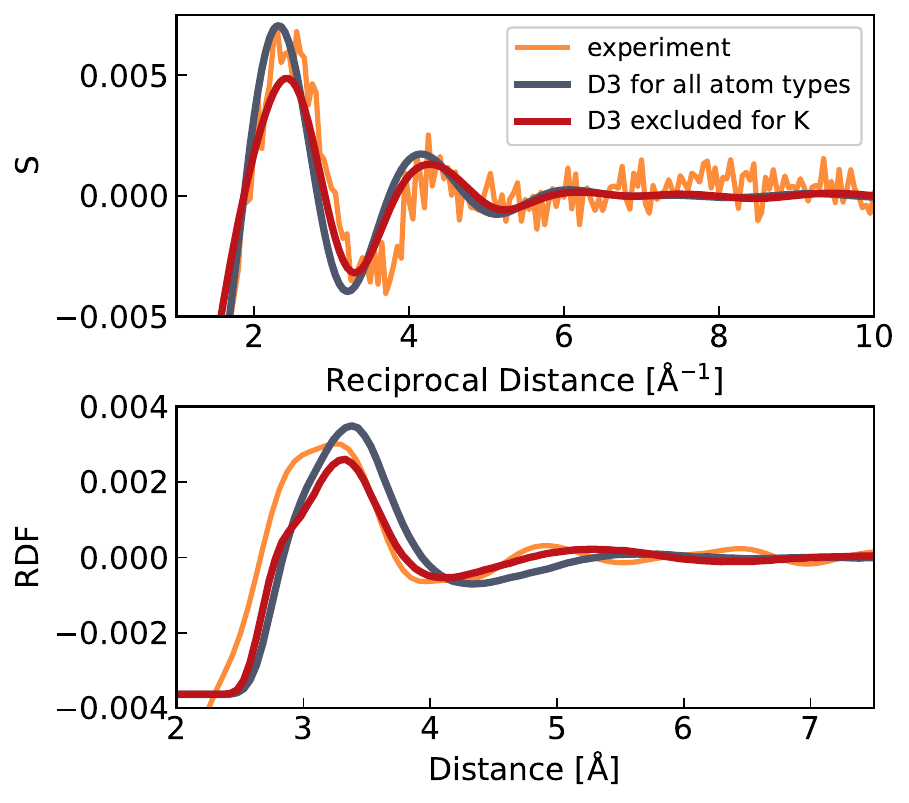}
    \caption
    {
    Simulated structure factor S (top) and radial distribution function (bottom) for potassium with water and chloride, shown in red (without D3 correction on potassium) and grey (with D3 correction on potassium).
    The experimental neutron scattering structure factor (top) and the corresponding RDF (bottom) from Reference~\citenum{Mason2019-10.1021/ACS.JPCB.9B08422} are depicted in orange for comparison.
    }
    \label{fig:rdf-K-r-q}
\end{figure}

Finally, we compared our simulated results for a 4~M potassium chloride solution to neutron scattering data~\cite{Mason2019-10.1021/ACS.JPCB.9B08422}.
The total radial distribution function (RDF) for potassium was obtained as a weighted sum of simulated RDFs for K--O, K--K, K--H, and K--Cl with appropriate weights from the reference \cite{Mason2019-10.1021/ACS.JPCB.9B08422}.
We then obtained its reciprocal space equivalent via Fourier transform in order to directly compare it with experimental neutron scattering data.
Our simulated RDFs, with and without the D3 correction for potassium, are shown along with the experimental curves in Figure \ref{fig:rdf-K-r-q} in both reciprocal and real space.
Clearly, better agreement with experiment was obtained without the D3 correction for potassium cation.

In this study, we used DFT-based AIMD simulations to examine the hydration structure of common alkali metal cations.
We found that the commonly used D3 correction scheme, which does not take into account the particular oxidation state, leads to inaccuracies in the hydration structure, as it assigns cations with dispersion coefficients appropriate for their neutral counterparts.
This results in highly polarizable and large cations which does not reflect experimental reality.
We evaluated this effect for cations including lithium, sodium, and potassium both at infinite dilution and for 4~M chloride solutions, which is particularly relevant when compared to the neutron scattering experiments.
To fix this issue, we propose to disable the D3 correction specifically for cation-including pairs, while still keeping the D3 term for solvent molecules (and anions). 
By comparison to neutron scattering data, we show that this simple correction significantly improved the hydration structures of the cations, in particular of sodium and potassium.
Therefore, we conclude that the commonly used D3 dispersion correction is often applied incorrectly when soft single-charged metallic cations are present.

As the computational power continues to increase, AIMD simulations of salt solutions are becoming a standard tool of computational chemistry.
These simulations offer good accuracy when condensed systems are addressed and may provide benchmark data for less accurate methods such as molecular dynamics with empirical potentials or data-driven methods, \textit{e.g.}, neural network potentials.
In this context, it is of key importance to properly account for dispersion interactions when studying alkali metal cations, which can be easily accomplished by zeroing the D3 term for all pairs involving a cation.

\section*{Computational Details}

The present simulations of bulk aqueous solutions were performed using the CP2K~9.1 package with the Quickstep module~\cite{Kuhne2020-10.1063/5.0007045} in the canonical ensemble at 300~K employing three-dimensional periodic boundary conditions.
Energy and forces were calculated on the fly using the revPBE~\cite{Perdew1996-10.1103/PhysRevLett.77.3865, Zhang1998-10.1103/PhysRevLett.80.890} density functional.
The dispersion interactions were accounted for using the D3 correction~\cite{Grimme2010-10.1063/1.3382344} scheme using two-body terms and zero damping and were either enabled or disabled for all pairs involving cations.
Using the hybrid Gaussian and plane wave approach~\cite{Lippert2010-10.1080/002689797170220}, Kohn--Sham orbitals were represented by a TZV2P basis set~\cite{VandeVondele2007-10.1063/1.2770708} and the charge density by an auxiliary plane-wave basis up to a 400~Ry cutoff.
The core electrons were replaced by Goedecker--Teter--Hutter (GTH) pseudopotentials~\cite{Goedecker1996-10.1103/PhysRevB.54.1703}.

To simulate each system using AIMD, we first pre-equilibrated it by means of force-field molecular dynamics (FFMD) in GROMACS~2022.2~\cite{Abraham2015-10.1016/J.SOFTX.2015.06.001} in the canonical ensemble for 10~ns with the SPC/E water and scaled-charge ion models (Reference~\citenum{Pluharova2014-10.1080/00268976.2013.875231, Kohagen2016-10.1021/ACS.JPCB.5B05221, Mason2012-10.1021/JP3008267, Kohagen2014-10.1021/JP5005693} for Li$^+$, Na$^+$, K$^+$ and Ca$^{2+}$ respectively).
Five initial structures were extracted from the FFMD trajectory with a 2~ns stride for each system which served as independent starting points for the subsequent AIMD simulations.
These five structures were then equilibrated for 5~ps using the Langevin thermostat~\cite{Allen2017} with a 50 fs time constant.
Next, the stochastic velocity rescaling (SVR) thermostat~\cite{Bussi2007-10.1063/1.2408420} was employed with a 1~ps time constant, and 20 and 40~ps of production runs were acquired for each structure.
This amounts to a total simulation time of 100 and 200~ps per system for the dilute and 4~M solution with chloride counterions.
Simulations with and without the cationic D3 correction were started from the same initial structures.

The dilute systems consisted of one cation surrounded by 128 water molecules in a cubic box.
The box size was chosen to match the density of pure liquid water of 1 kg/dm$^3$ resulting in 15.656~\AA\ for lithium, 15.692~\AA\ for sodium, 15.728~\AA\ for potassium and 15.730~\AA\ for calcium cation.
The 4~M solutions contained 8 cations, 8 chloride anions, and 111 water molecules.
The box sizes for these systems, as determined based on the experimental densities of the salts, were 15.314~\AA\ for LiCl~\cite{advancedthermo_density_licl}, 15.328~\AA\ for NaCl~\cite{CRCHandbook-2016-nacl-aq-dens} and 15.524~\AA\ for KCl~\cite{advancedthermo_density_kcl}.

Free energy profiles of ion pairing ($A$) were calculated from the RDFs in a standard manner as follows
\begin{equation}
    \label{eq:free-energy}
    A = - k_\mathrm{B}T\log(\mathrm{RDF})
\end{equation}
where $k_B$ is the Boltzmann constant and $T$ temperature.

Finally, gas-phase calculations were performed using the ORCA~5.0.3 software~\cite{Neese2020-10.1063/5.0004608, Neese2022-10.1002/WCMS.1606}.
The interaction energy ($E_\mathrm{int}$) curve was obtained as an energy scan of the distance between the cation and the oxygen of the water molecule or another cation, as described by the equation:
\begin{equation}
    E_\mathrm{int}=E_\mathrm{MX}^\mathrm{MX} - E_\mathrm{M}^\mathrm{MX} - E_\mathrm{X}^\mathrm{MX}
\end{equation}
where the subscript denotes the system (M: metal cation, X: water or metal cation), and the superscript indicates the basis set used.
Such $E_\mathrm{int}$ calculation includes a counterpoise correction for the basis set superposition error~\cite{Boys2006-10.1080/00268977000101561}.
The revPBE~\cite{Perdew1996-10.1103/PhysRevLett.77.3865, Zhang1998-10.1103/PhysRevLett.80.890} density functional (optionally equipped with the D3~\cite{Grimme2010-10.1063/1.3382344} or D4~\cite{Caldeweyher2017-10.1063/1.4993215, Caldeweyher2019-10.1063/1.5090222} dispersion correction) and coupled cluster single, double and perturbative triple excitations [CCSD(T)]~\cite{Bartlett2007-10.1103/REVMODPHYS.79.291} methods were used for the $E_\mathrm{int}$ calculation.
In all the gas-phase calculations, the def2-TZVPP~\cite{Weigend2005-10.1039/B508541A} basis set was employed.

\section*{Supporting Information}

The isolated D3 contribution to the interaction energy as a function of distance together with a comparison to the D4 correction.
RDF and RCN for calcium at infinite dilution.
RDFs of water in all the systems, as well as additional oxygen--anion RDFs in concentrated solutions.

\section*{Acknowledgements}

The authors thank Stefan Grimme for valuable comments.
P.J and H.M.-S. acknowledge the support from the Czech Science Foundation
(EXPRO Grant 19-26854X).
V.K. acknowledges Faculty of Science of Charles University where he is enrolled as a PhD. student and the IMPRS for Many Particle Systems in Structured Environments Dresden.
We acknowledge ChatGPT for improving readability and language.

\end{bibunit}

\clearpage

\setcounter{section}{0}
\setcounter{equation}{0}
\setcounter{figure}{0}
\setcounter{table}{0}
\setcounter{page}{1}

\renewcommand{\thesection}{S\arabic{section}}
\renewcommand{\theequation}{S\arabic{equation}}
\renewcommand{\thefigure}{S\arabic{figure}}
\renewcommand{\thetable}{S\arabic{table}}
\renewcommand{\thepage}{S\arabic{page}}
\renewcommand{\citenumfont}[1]{S#1}
\renewcommand{\bibnumfmt}[1]{$^{\rm{S#1}}$}

\title{Supporting information for: \mstitle}
{\maketitle}

\begin{bibunit}

\section{Contributions of the D3 and D4}

\begin{figure}[b!]
    \centering
    \includegraphics[width=\linewidth]{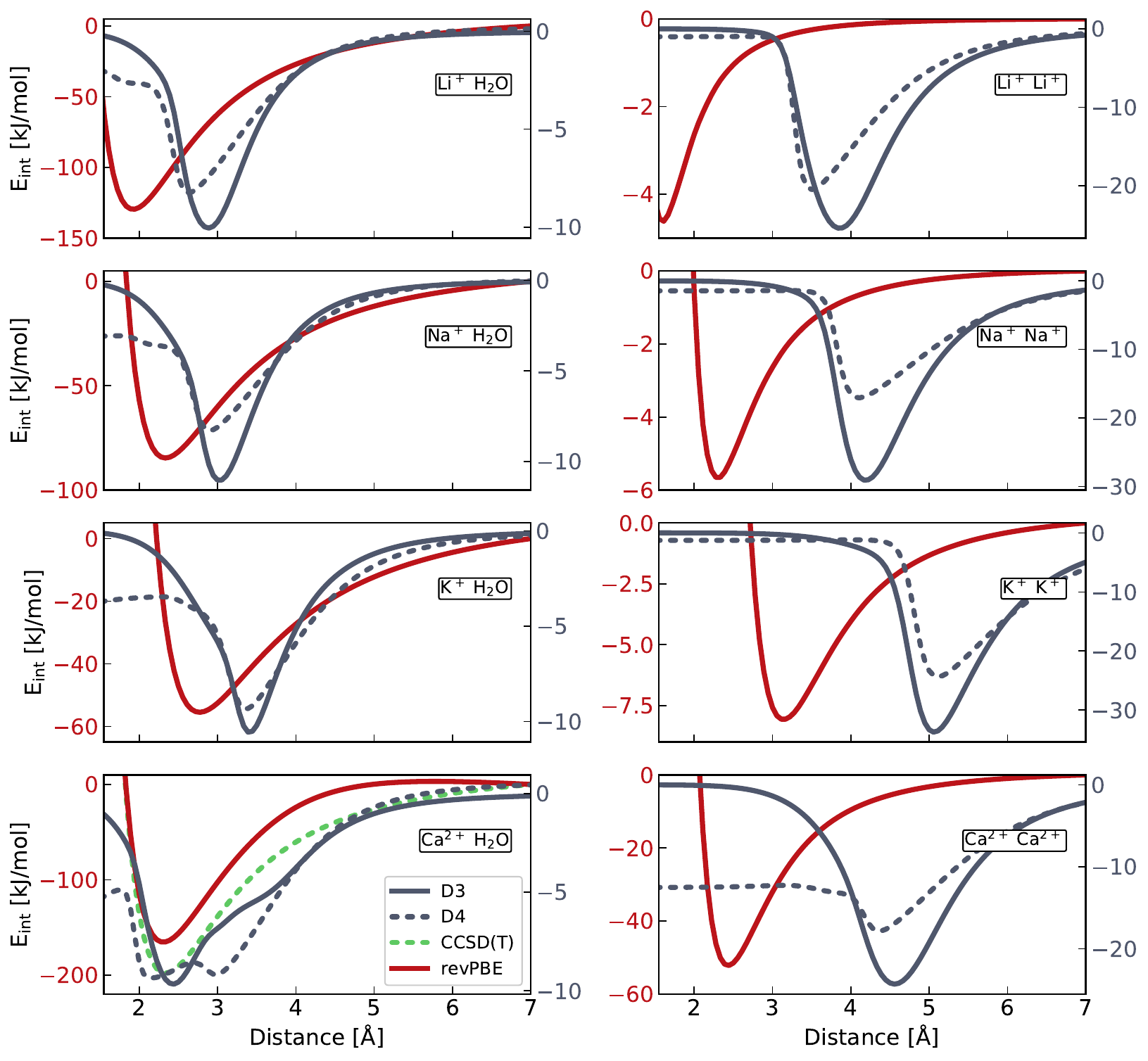}
    \caption
    {
    $E_\mathrm{int}$ of cation--water (left) and cation--cation (right) as a function of the distance for lithium, sodium, potassium, and calcium cations from top to bottom.
    $E_\mathrm{int}$ obtained at revPBE level is plotted on the left $y$-axis in red and the D3 (solid) and D4 (dashed) dispersion correction on the right $y$-axis in grey.
    Additional CCSD(T) (dashed green line) result is displayed for calcium cation.  
    Note that Coulomb interaction potential was subtracted from the cation--cation curves.
    }
    \label{fig:D3-D4}
\end{figure}

In Figure~\ref{fig:D3-D4}, we show gas-phase interaction energies of a cation with water (left panel) and the same cation (right) obtained according to the Equation~2 in the main text for Li$^+$, Na$^+$, K$^+$, and Ca$^{2+}$.
$E_\mathrm{int}$ calculated by the revPBE density functional and the CCSD(T) (in the case of calcium) method are compared on the left-hand side $y$-axis, while the isolated D3 and D4 dispersion correction to the revPBE energies on the right-hand side $y$-axis.
It is worth a note that we subtracted Coulomb potential from the cation--cation curves on the right-hand side panels of Figure~\ref{fig:D3-D4} in order to facilitate the readout of features that would be otherwise covered by the charge--charge repulsion.

The D3 contributes to the $E_\mathrm{int}$ at distances larger than where the minimum is located, which agrees with the results presented in Figure~1 of the main text.
This contribution amounts roughly to 10~kJ/mol for cation--water curves.
For the cation--cation cases, it increases to 20--30~kJ/mol.
The magnitude of the D4 dispersion correction is generally smaller than that of D3.
However, the problematic position of the peak is practically the same for both corrections in spite of D4 reflecting better varying electronic structure of the involved species with respect to D3.
In the case of calcium, used here as an example of a divalent cation, we see a significant discrepancy between $E_\mathrm{int}$ calculated either with revPBE and the CCSD(T).
The difference is a consequence of the spurious overdelocalization of electrons in the gas phase, which is not surprising for the GGA DFT. This artifact does not occur at the CCSD(T) level.

Additional RDFs of water and chloride anion are shown in this section for the infinitely diluted and concentrated systems in Figure~\ref{fig:rdfs-water}.
It reveals that neither structure of water nor the hydration of chloride anion is perturbed by the exclusion of metal cation from the D3 dispersion calculation.

\section{Calcium cation at infinite dilution}

\begin{figure}[t!]
    \centering
    \includegraphics[width=\linewidth]{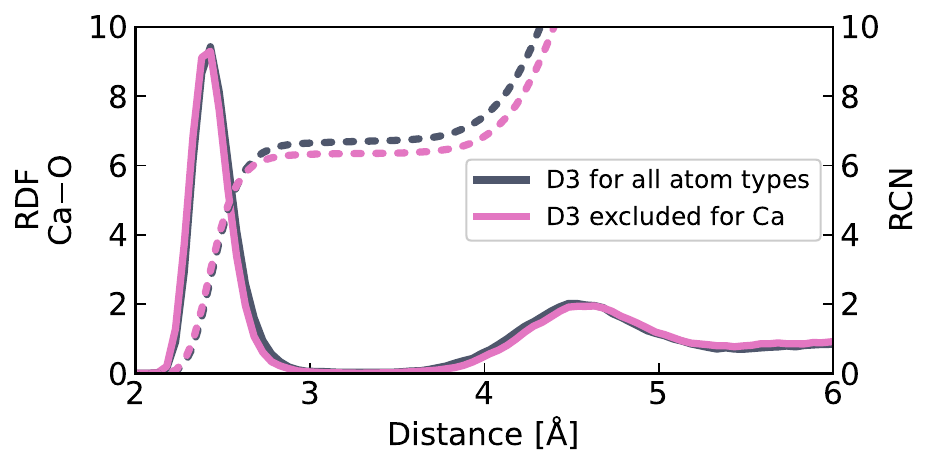}
    \caption
    {
    RDFs of calcium--oxygen (full) and corresponding RCN
    (dashed) including the D3 correction for all atom (grey) and for
    all kinds but calcium (pink).
    }
    \label{fig:rdf-rcn-Ca}
\end{figure}

We simulated a single calcium cation at infinite dilution according to the protocol in the Computational Details of the main text.
We extracted the Ca--O RDF and RCN with and without the D3 correction applied to the calcium cation.
Both setups yielded similar results, with the first peak at 2.43~\AA.
This agrees well with the value of 2.38~\AA\ obtained in the neutron scattering experiment~\cite{Kohagen2014-10.1021/JP5005693}.
The main difference between the two simulations is the average CN.
In the case where the D3 correction was applied to the calcium cation, the CN is slightly larger at 6.68 water molecules in the first hydration shell.
In contrast, a CN of 6.33 is obtained when the cationic D3 correction is not used.
This observation is consistent with the results presented for lithium, sodium, and potassium in the main text.

\section{water structure}

\begin{figure}[h!]
    \centering
    \includegraphics[width=\linewidth]{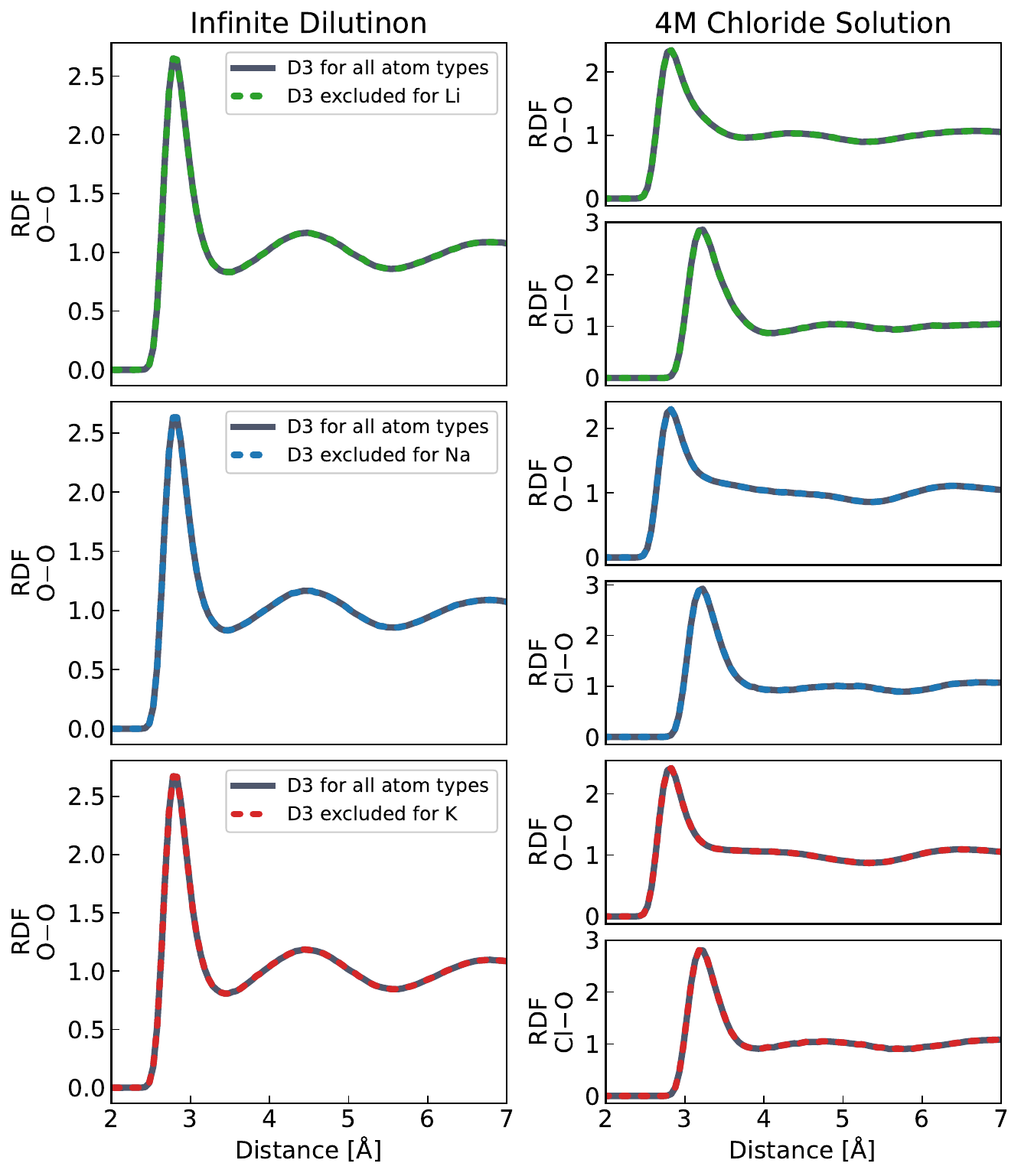}
    \caption
    {
    RDFs in the infinitely diluted system (left panels) and in the 4M chloride solution (right panels).
    Oxygen--oxygen RDFs are displayed for all systems, and oxygen--chloride RDFs only for the 4~M solutions.
    Cations are color-coded as green (lithium), blue (sodium), and red (potassium).
    }
    \label{fig:rdfs-water}
\end{figure}

\end{bibunit}

\end{document}